\documentclass[aps,prl,twocolumn,floatfix,superscriptaddress]{revtex4}
\pdfoutput=1
\usepackage{latexsym}       
\usepackage{graphicx}       
\usepackage{rotating}       
\usepackage{hyperref}
\usepackage{amsmath,amssymb,amsfonts}
\usepackage{bm}
\usepackage{graphicx}
\usepackage{color}

\newcommand{\figwidth}{0.72\columnwidth}

\newcommand{\dx}{d$_{x2-y2}$ }
\newcommand{\psig}{p$_\sigma$ }

\newcommand{\lasco}{La$_{2}$CuO$_4$ }

\newcommand{\lascoSr}{La$_{2-x}$Sr$_x$CuO$_4$ }

\newcommand{\emphasize}{\emph}

\def\onlinecite#1{\cite{#1}}

\newcommand{\up}{\uparrow}
\newcommand{\dn}{\downarrow}

\newcommand{\vk}{\bold{k}}

\begin{document}

\title{Optical Weights and Waterfalls in  Doped Charge Transfer
  Insulators: an LDA+DMFTStudy of LSCO }
\author{C\'edric Weber}
\author{Kristjan Haule}
\author{Gabriel Kotliar}
\affiliation{Department of Physics, Rutgers University,  Piscataway,
  NJ 08854, USA}


 \begin{abstract}
   We use the Local Density Approximation in combination with the
   Dynamical Mean Field Theory to investigate intermediate energy
   properties of the copper oxides.  We identify coherent and
   incoherent spectral features that results from doping a charge
   transfer insulator, namely quasiparticles, Zhang-Rice singlet band,
   and the upper and lower Hubbard bands.  Angle resolving these
   features, we identify a \emphasize{waterfall} like feature, between
   the quasiparticle part and the incoherent part of the Zhang-Rice
   band.  We investigate the assymetry between particle and hole
   doping. On the hole doped side, there is a very rapid transfer of
   spectral weight upon doping in the one particle spectra. The
   optical spectral weight increases superlinearly on the hole doped
   side in agreement with experiments.
 \end{abstract}

 \maketitle


Since their discovery, the high temperature superconductors continue
to be a subject of intensive investigations. It is widely believed
that strong correlations and many body effects are responsible for
many of the peculiar properties of these materials. However, after
many years of intensive studies, a comprehensive understanding of
their electronic structure, even at intermediate energy scales, is
lacking. The question of wether the strengh of the on site
correlations is sufficient to open the gap at half filling (Mott
picture) or alternatively magnetism is required (Slater picture) is
actively debated \cite{luca_nature}. A vertical feature in the
photoemission spectral intensity, commonly known as a
waterfall, has recently been found in many compounds \cite{valla_kink,Inosov_watefalls},
and it has been assigned to several conflicting microsopic origins
\cite{savrasov_waterfall,macridin_kink,zemlji_kink}.

It is generally accepted that these materials contain copper-oxygen
layers as a common element. Stoichiometric compounds such as \lasco,
are antiferromagnetically ordered charge transfer insulators in the
Zaanen-Sawatsky Allen \cite{ZSA_theory_Sawatzky,ZSA_theory_Sawatzky_2}
classification scheme. Hence a minimal model of these materials
contains two oxygen and one copper orbital per unit cell
\cite{emery_3bands_model,varma_3bands_model}. When this model is doped
with electrons, the low energy physics is believed to be closer to that
of the Hubbard model. On the hole doped side, this reduction
constructed by Zhang and Rice \cite{zhang_rice_singlet,zhang_singlet}
can be achieved in the limit that the charge transfer energy is much
larger than the copper oxygen hybridization. However, there is no
agreement on the precise region ( in energy, and parameter space )
where the reduction to a one band model is accurate.

In this letter, we use a realistic theoretical approach, i.e., the
Local Density Approximation combined with the Dynamical Mean Field Theory
(LDA+DMFT) \cite{our_review}, to study a typical cuprate compound
\lascoSr.  The LDA calculation was done by PWSCF package
\cite{cappuccion}, which employs a plane-wave basis set and ultrasoft
pseudopotentials \cite{lda_vanderbilt}.  Downfolding to a three band
model, containing copper \dx and two oxygen \psig orbitals was
performed by the maximally localized Wannier functions (MLWF) method
\cite{lda_basis,lda_basis2}.  We solve this model using Dynamical Mean
Field Theory with an exact impurity solver
\cite{Haule_DMFT}.

\begin{figure}
\begin{center}
\includegraphics[width=\figwidth]{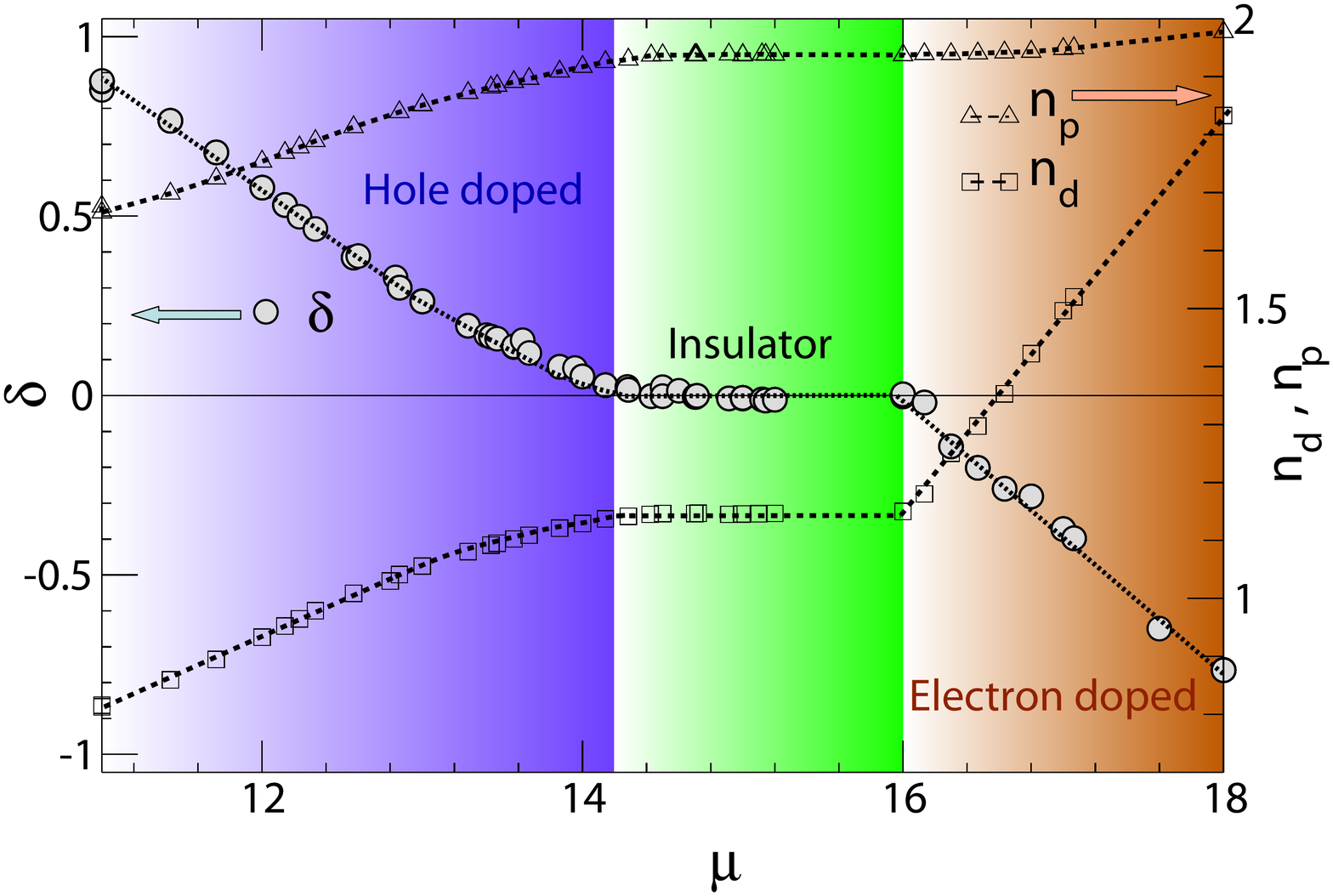}
\caption{(Colors online) 
  Doping $\delta$ (left scale) versus the chemical potential $\mu$.
  The parent compound ($\delta=0$) is a charge transfer insulator with
  a gap $\Delta_{pd}\approx 1.8eV$ (green region).  The number of electrons in the
  \dx orbital ($n_d$) and in the \psig orbitals ($n_p$) are also shown (right
  scale).}
\label{fig:phases}
\end{center}
\end{figure}

We address the following questions. What are the spectral features
that result from doping a charge transfer insulator?  How are these
spectral features distributed in momentum space?  How does the
quasiparticle residue and the optical weight, induced by doping the
charge transfer insulator, vary with doping?  We address the
similarities and differences between our results and those that follow
from a Hubbard model description.

Downfolding the LDA band structure of \lasco results in the following three band Hamiltonian:
\begin{multline}
  \label{eq:3band_hub1}
  \mathcal{H}_t = \sum\limits_{ij\sigma, (\alpha,\beta) \in (p_x,p_y,d_{x2-y2})}{  t^{\alpha \beta}_{ij} c^\dagger_{i \alpha \sigma} c_{j \beta \sigma} } \\ 
                + \epsilon_p \sum_{i \sigma \alpha \in (p_x,p_y)}{\hat n_{i \alpha \sigma}} + 
                  \left(\epsilon_d-E^{dc}\right) \sum_{i\sigma}{\hat n_{i d \sigma}}
\end{multline}
where $i$ and $j$ label the CuO$_2$ unit cells of the lattice, and
$t_{ij}^{\alpha\beta}$ are the hopping matrix elements.
$\epsilon_d$ and $\epsilon_p$ are the on-site energies of the $d$ and
$p$ orbitals, respectively. 

The charge transfer energy, $\Delta_{pd}$, between the copper and oxygen
plays the role of an effective onsite repulsion $U$ in a Hubbard model
picture, as seen for example in slave bosons mean-field studies
\cite{review_gabi}.  

The LDA downfolding procedure results in $\epsilon_d-\epsilon_p=2.78eV$.
To this hamiltonian, we add the onsite Coulomb repulsion $U$ on the
\dx orbital
\begin{equation}
\label{eq:3band_hub2}
\mathcal{H}_U = U_d \sum_{i}{ \hat n_{id\up} \hat n_{id\dn}} 
\end{equation}
where the value of $U_d=8 eV$.  The LDA+DMFT method, accounts for the
correlations which are included in both LDA and DMFT by a double
counting correction to the $d$-orbital energy, which we take to be
fixed at $E_{dc}=3.12eV$ for all dopings.  We neglect the oxygen
Coulomb repulsion $U_p$.  Within single site DMFT, the copper-oxygen
repulsion $V_{dp}$ is treated at the Hartree level, and hence it is
included in the orbital energies of the $p$ and the $d$ orbitals.

The Green function of the three band model is given by:
 \begin{equation}
   \label{greenfunc}
   \textbf{G}_\vk(i \omega_n) =  \left( i \omega_n + \mu - \bold{H}_\vk - \bold{\Sigma}(i \omega_n)   \right)^{-1},
 \end{equation}
where $\bold{H}_\vk$ is the Fourier transform of the $\mathcal{H}_t$ in
Eq.~(\ref{eq:3band_hub1}) and is a $3\times 3$ matrix. $\bold{\Sigma}$ is the self-energy
matrix being nonzero only in the $d$ orbital.

\begin{figure}
\begin{center}
\includegraphics[width=\figwidth]{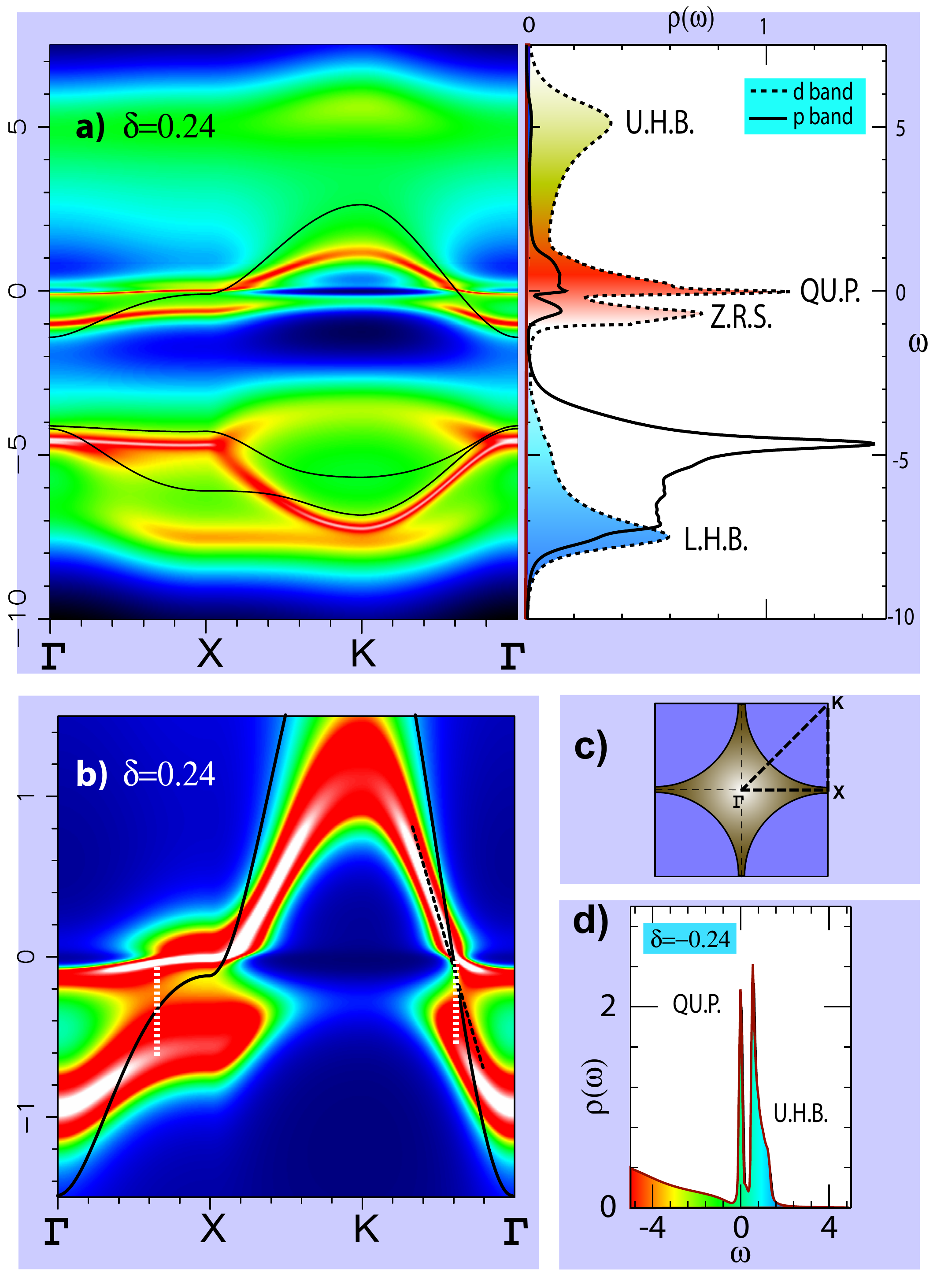}
\caption{(Color online) 
  a) Left panel: Spectral function $A(\vk,\omega)$ for the hole doped
  compound ($\delta=0.24$). The black lines correspond to the LDA
  bands.  Right panel: The local spectral functions of the \dx band
  (dashed line) and of the \psig band (full line). We marked the
  positions of the lower Hubbard band (L.H.B.), coherent part of the
  Zhang-Rice singlet called the quasi-particle peak (QU.P), the
  incoherent part of the Zhang-Rice singlet band (Z.R.S.)  and the
  upper Hubbard band (U.H.B.).  b) The low energy part of $A(\vk,\omega)$
  from panel (a). The dashed line shows the slope of the quasiparticle
  band, which gives the Fermi velocity. The derivative was obtained by
  a quadratic expansion of $\Sigma(\omega)$ at the Fermi level.  The
  Fermi velocities in LDA+DMFT is $v_{DMFT}=2.15(5)\,\textrm{\AA eV}$ in very
  favorable agreement with experiment ($v_{exp}=2.0\,\textrm{\AA eV}$ at doping
  $\delta=0.24$ \cite{Zhou_nature_paper}). The ratio between the DMFT
  and the band velocity is thus $v_{DMFT}/v_{LDA}=0.49$.  c) Fermi
  surface and the path along which $A(\vk,\omega)$ is shown in (a) and
  (b) 
  d) The density of states in the electron doped compound.  A very
  sharp quasi-particle peak develops close to the upper Hubbard band
  in the electron doped case.  }
\label{fig:holedoped}
\end{center}
\end{figure}
The self energy in Eq.~(\ref{greenfunc}) is obtained
by solving an Anderson impurity model subject to the  DMFT
self-consistency condition:
\begin{equation}
 \label{greenfunc}
 \frac{1}{ i\omega - E_{imp} - \Sigma(i\omega) - \Delta(i\omega)}  = \frac{1}{N_k} \sum\limits_{\vk \in BZ}  {G_\vk^{dd}(i \omega)},
\end{equation}
where the sum runs on the first Brillouin Zone (BZ).  In this work we
use the continuous time quantum Monte Carlo impurity solver algorithm
\cite{Haule_DMFT,werner_ctqmc_algorithm}, which gives the self-energy
functional $\Sigma [ E_{imp}, \Delta ]$ within the residual
statistical error bars of the Monte Carlo algorithm
\footnote{In this work we used temperature $T=116K$.}
Real frequency resolved quantities were obtained by analytic
continuation (more details will be given elsewhere
\cite{Haule_longer_paper_many_band}) of the observables on the
imaginary axis. We have crossed checked the analytic continuation
using the OCA real frequency solver \cite{Haule_new}.

Equations (\ref{eq:3band_hub1}) and (\ref{eq:3band_hub2}) were studied
previously in
Refs.~\onlinecite{dmft_paper_krauth_gabi,Avignon_DMFT_perovskite}.
When $U_d$ is large, there is a metal to charge transfer insulator
transition at integer filling, as a function of the charge transfer
energy $\epsilon_d-\epsilon_p$.
For the set of realistic parameters considered in this letter, we find
that the parent compound is a charge transfer insulator (see Fig.
\ref{fig:phases}), with a hole density per Cu atom $\approx 85\%$ and
a charge gap close to $1.8 eV$. This places this material slightly
above (but not very far) from the metal to charge transfer insulator
transition point.
This is demonstrated in Fig. \ref{fig:phases}, where the number of particles ($\delta$) as 
a function of chemical potential ($\mu$) exhibits a plateau in the interval $\mu=14.2eV{-}16eV$.
This figure also display the partial occupancies, describing the relative
distribution of the occupancies among copper and oxygen.  
\begin{figure}
\begin{center}
\includegraphics[width=\figwidth]{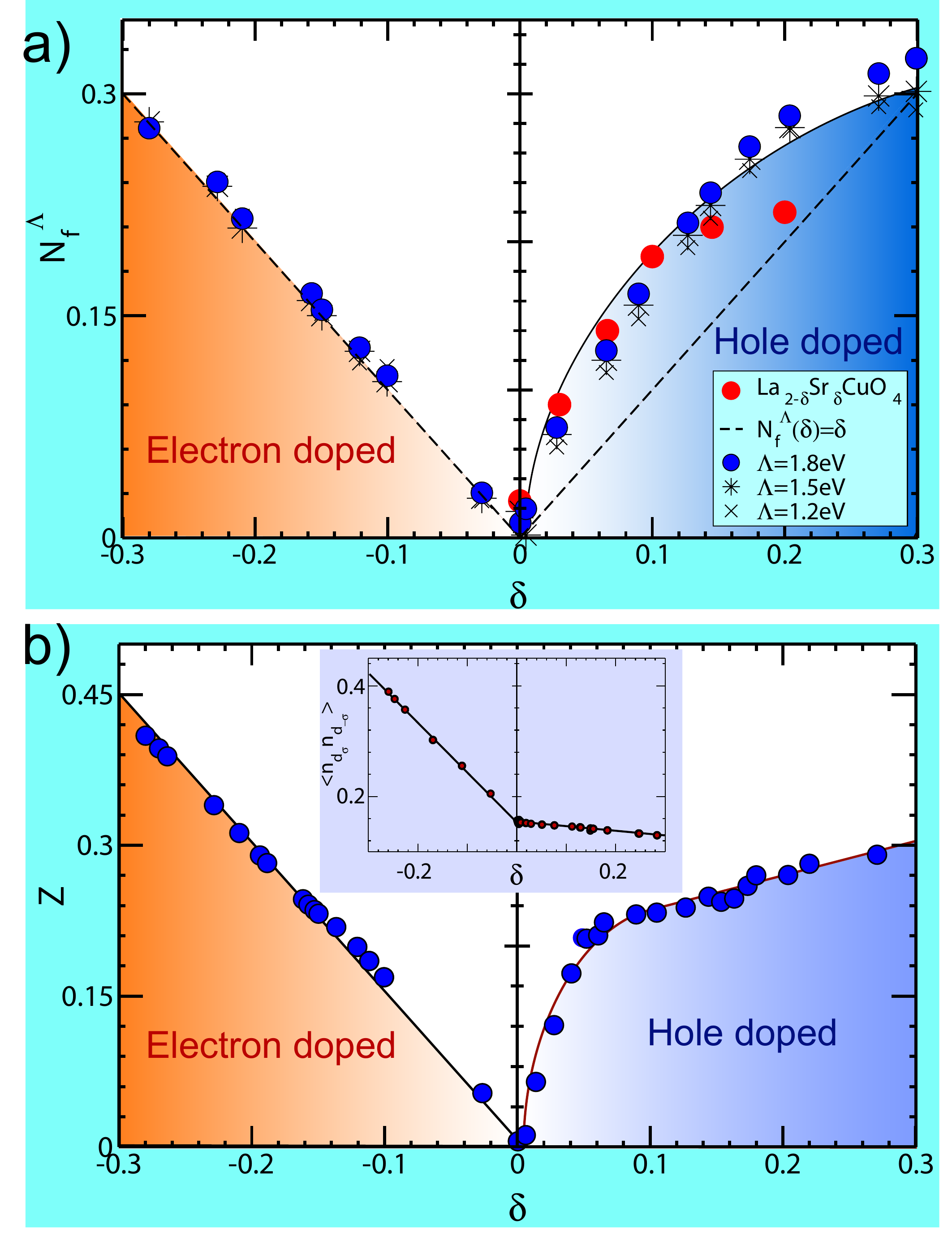}
\caption{(Color online) 
  a) Optical spectral weight $N_{eff}$ obtained by Eq.~\ref{Sweight}
  using various cutoffs $\Lambda=1.2,1.5,1.8eV$ for both the hole
  doping ($\delta>0$) and electron doping ($\delta<0$).
  The experimental data (red circles) are reproduced from
  Ref.~\onlinecite{experimental_Nf}.
  Long dashed line corresponds to the curve $N_{eff}(\delta)=\delta$.
  Note the assymetry between the electron ($\delta<0$) and hole doped
  ($\delta>0$) side.  
  b) The quasi-particle weight $Z$ versus doping.  
  Inset: double occupancy of the \dx orbital versus doping. 
  For the electron doped compound, the double occupancy increases
  linearly with doping, whereas the hole doping has only a weak
  effect on double occupancy of the \dx orbital.  
  Continuous lines are guide to the eyes.
}
\label{fig:doping}
\end{center}
\end{figure}

\begin{figure}
\begin{center}
\includegraphics[width=\figwidth]{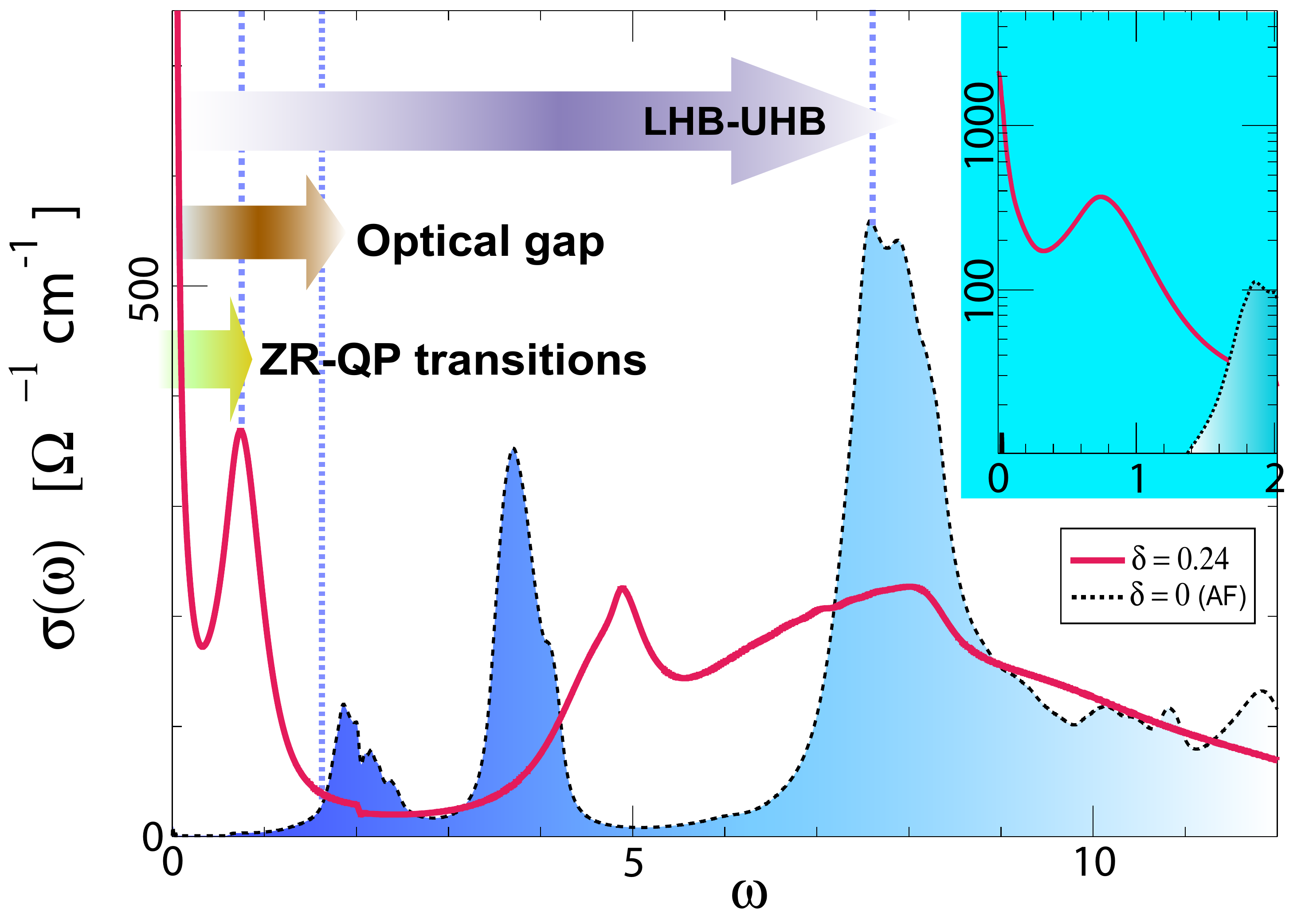}
\caption{(Color online) Optical conductivity $\sigma(\omega)$ for the
  parent compound ($\delta=0$) and the hole doped compound
  ($\delta=0.24$).  For the parent compund, we show optics in the Neel
  state. The optical gap in the parent compound is around 1.8$\,$eV in
  both the paramagnetic and antiferromagnetic state.  The structure
  around $\omega=8eV$ corresponds to transitions between the lower and
  the upper Hubbard band.  Upon hole doping, additional peak appears
  close to $\omega \approx 1eV$ for a large range of doping
  $\delta=5-30\%$. It corresponds to transitions between the
  quasi-particle part and incoherent part of the Zhang-Rice singlet.}
\label{fig:optics}
\end{center}
\end{figure}
We have computed the spectral functions for both the parent
and a doped compound. In the parent compound (not shown), we found
four spectral peaks: i) the upper Hubbard band around +2.5 eV, ii) a
singlet bound state of oxygen and copper known as the Zhang-Rice
singlet just below the Fermi level, iii) the oxygen set of bands
centered around -7eV and iv) the lower Hubbard band at -8eV.  By hole
doping (see Fig. \ref{fig:holedoped}a-right) , the Zhang-Rice band is split
into two peaks, the quasi-particle peak and the incoherent part which
is associated with the lower Hubbard band in a single-band Hubbard
model.  

Figures \ref{fig:holedoped}a-left, and \ref{fig:holedoped}b show the
momentum resolution of these features. The quasiparticle band crosses
the Fermi level with a Fermi velocity along the nodal direction of the
order of $v_{Fermi}=2.2 \textrm{\AA eV}$ in good agreements with
experimental results \cite{Zhou_nature_paper}. Figure
\ref{fig:holedoped}b also shows a spectral gap around -0.3$\,$eV,
separating the coherent and incoherent parts of the Zhang Rice band,
with weak spectral weight connecting them. This is reminiscent of the
waterfall feature observed in numerous cuprates.  Moreover, our
calculations lead to a very good agreement with experiments for both
the range of energy where the high energy peak is present ($\approx
-1eV$) and the position in the Brillouin zone where the abrubt change
is observerd (close to $(\pi/2,\pi/2)$).  While there are many
interpretations of this feature
\cite{savrasov_waterfall,macridin_kink,zemlji_kink,Inosov_watefalls,valla_kink},
we find that, a proper description of the waterfalls requires a
multi-band model containing both oxygen and copper orbitals, as
pointed out in Ref.~\cite{savrasov_waterfall}.  The waterfall comes
from the spectral gap between the incoherent part and the coherent
(quasiparticle) part of the Zhang-Rice band.
The latter is not present in the single site DMFT description of the
parent compound, and our calculation in paramagnetic state of \lasco
does not show splitting of the Zhang-Rice band into a low energy and a
high energy part.
We have found however, that when the anti-ferromagnetic broken
symmetry state is considered for a parent compound, the splitting of
the Zhang-Rice singlet is present, and the low energy part of the
Zhang-Rice band is touching the Fermi level at $(\pi/2,\pi/2)$
momentum.

One of the most notable aspects of cuprate physics, is the rapid
growth of the optical conductivity inside charge gap upon hole doping
\cite{experimental_Nf}. This was noted early on by Sawatzky and
collaborators \cite{Sawatzky_spec_weight}. In order to tackle this
issue, we computed the optical conductivity, given by:
\begin{multline}
 \sigma'(\omega) = \sum\limits_{\sigma \vk}{\frac{e^2}{\hbar c \pi}  \int{ dx \frac{  f(x-\omega)-f(x)}{\omega}}} 
                  \\ \times \textrm{Tr}\big( \hat{\bold{\rho}}_{\vk\sigma} (x - \omega)  \bold{v}_\vk  \hat{\bold{\rho}}_{\vk\sigma}(x) \bold{v}_\vk \big)
\end{multline}
Where $c$ is the interlayer distance, the density matrix $\bold{\hat{\rho}}$ is defined by
\begin{equation} 
  \hat{\bold{\rho}}_{\vk\sigma}(x)=\frac{1}{2\pi i} 
\left(\bold{G}^\dagger_{\vk\sigma}(x)-\bold{G}_{\vk\sigma}(x)\right)
\end{equation}
And the bare vertex is given by $\bold{v_k}={d\bold{\hat H}_\vk}/{d k_x}$.  
Our results are dispayed in Fig.~\ref{fig:optics}.  
The optical gap in the parent compound is found to be $1.8 eV$. 
To quantify the rate of the redistribution of optical spectral weight, we computed the effective electron
number per Cu atom defined by
\begin{equation}
N^{\Lambda}_{eff}=\frac{2 m_e V}{\hbar \pi e^2} \int^{\Lambda}_{0}{\sigma'(\omega) d\omega},
\label{Sweight}
\end{equation}
where $m_e$ is the free electorn mass, and $V$ is the cell Volume
containing one formula unit. $N_{eff}$ is proportional to 
the number of electrons involved in the optical excitations up to the cutoff
$\Lambda$. Our results for $N_{eff}$ are displayed in  Fig.~\ref{fig:doping} and
compared to experimental data taken from Ref.~\cite{experimental_Nf}.

Notice that the spectral weight, contained in the region below the
charge transfer gap of the parent compound, grows in a superlinear
fashion. This is in agreement with experiments of
Ref.~\cite{experimental_Nf} but is incompatible within a rigid band
picture of doping either a band or a Mott insulator far from the Mott
transition (as for example in the Hubbard model well above $U_{c2}$).
Thus the optical spectral weight in the charge transfer insulator grows 
faster with doping than in the Mott insulator, where the linear growth of the spectral weight is expected.

The realistic three-band model within single site DMFT approach leads
to a considerable asymmetry in the electron and hole doped side of the
phase diagram.  The superlinear behavior found in our calculation of
the three band model is sufficient to explain the experimental
findings, even though the correlations are strong enough to open a
charge transfer gap in the parent compound even in the paramagnetic state.

We verified that this interesting particle hole assymetry is
present also in the quasiparticle residue $Z=\left(1 - \frac{ d
    \Sigma'(\omega)}{ d \omega } \right)^{-1}$ and is shown in
Fig.~\ref{fig:doping}b.
This is a one electron quantity that can be obtained directly from the
imaginary axis data without the need to invoke analytic continaution,
but still follows the same trend as $N_{eff}$.

It is interesting to compare the asymmetry found here with the recent
results of Ref.~\cite{marcello_mit_anderson} on the Anderson lattice model.
The authors found a much larger particle hole assymmetry compared to
our results. Most likely this is due to a momentum independent
hybridization function, which makes the Kondo scale in the hole doped
side very small \cite{meanfield_cuprates_millis_kotliar}, and hard to
reach with conventional QMC.

Phenomena in the underdoped region, such as the normal state pseudogap
and the formation of fermi arcs, are not described by single site DMFT
and require cluster extensions of the formalism \cite{Haule_long_paper_CTQMC}.
However, single site DMFT is still expected to describe well
intermediate energy phenomena, such as the distribution of the
spectral weight upon doping the Mott/charge transfer insulator, which
is the focus of our work.  Momentum resolved quantities become more
accurate in the overdoped region, where the self energy is more local.

For this reason, we have not considered here the details of the
momentum dependent photoemission spectra at low doping, or the details
of the frequency dependent optical conductivity, which are
considereably influenced by cluster effects. We have rather focused on
quantities such as $N_{eff}$, which are less sensitive to cluster
corrections and we rather presented the detailed photoemission spectra
in the overdoped region, where actually the local approximation is
valid \cite{Haule_long_paper_CTQMC}.

In conclusion, we have carried out a realistic LDA+DMFT calculations
accounting for the \dx and \psig orbitals of the three band model.
The electron doped side is behaving mainly like the single-band
Hubbard model. On the other hand, the hole doped side of the phase
diagram is very different. a) We find a superlinear optical spectral
weight transfer in agreement with experiments. b) On the spectral
function, we find a spectral gap between the quasiparticle band and
the incoherent Zhang Rice singlet band. This leads to a very abrupt
and nearly vertical change in the spectral intensity in the nodal
direction of the spectrum (\emph{waterfall}) , that ends up below
1$\,$eV.

Our LDA+DMFT study support the view that the hole doped cuprates are
above, but not very far from the metal charge transfer insulator
transition in agreement with the conclusion of earlier slave boson
studies \cite{review_gabi}.

It would be interesting to apply the same methodology to the electron
doped cuprates which have a smaller charge transfer gap. These
materials might have a charge transfer energy which lies below the
critical value needed to sustain a paramagnetic insulator at integer
filling, and hence might undergo a different metallization process
upon doping, as suggested in Ref.~\cite{senechal_tremblay}.

Acknowledgement: We thank 
C.A. Marianetti, L. de' Medici and M. Ferrero and C. Varma for many illuminating
discussions.  
C.W. and G.K. were supported by NSF grant No. DMR 0528969.


\bibliographystyle{prsty}

\end{document}